\documentclass[11pt,a4paper]{article}
\usepackage{subfigure}
\usepackage{xcolor}
\usepackage[T1]{fontenc} % if needed
\usepackage{xspace}
\usepackage{multirow}
\usepackage{hyperref}
\usepackage{slashed}

\usepackage{graphicx,amsmath}
 \usepackage{amssymb}
\usepackage{lineno,color}
\usepackage{url}

\newcommand{\p}{I\!\!P}
\def\be{\begin{equation}}                                                    
\def\ee{\end{equation}}                                                    
\begin{document}
%\numberwithin{equation}{section}
% \linenumbers    
 
%\rightline{IPPP/21/41}

\bigskip
\bigskip
\begin{center}
{\bf\Large Large Rapidity Gaps in proton-nucleus interaction. }\\
 
 \vspace{1cm}
   
 V.~A.~Khoze$^{a}$ and 
 M.~G.~Ryskin$^{b}$\\
 
 \vspace{0.7cm}
 $^a$ Department of Physics, Institute for Particle Physics Phenomenology,\\  Durham University, Durham, DH1 3LE, UK\\
 $^b$ Petersburg Nuclear Physics Institute, NRC Kurchatov Institute, \\Gatchina, St.~Petersburg, 188300, Russia\\
\
% $^b$ Department of Physics, Institute for Particle Physics Phenomenology,\\  Durham University, Durham, DH1 3LE, UK\\
%}
 \vspace{0.7cm}

 \abstract{\noindent We analyse the cross-section of events with Large Rapidity Gaps observed 
 in proton-lead collisions by the CMS collaboration\cite{CMS-Pb}. The role of the transverse size of elementary $pN$ amplitude is discussed. We emphasize that the cross-section of incoming proton dissociation
  caused by the photon radiated off the lead ion is close to the value of $d\sigma/d\Delta\eta^F$  
  measured by the CMS, and it is not clear why there is no room in the data 
for the Pomeron-induced contribution }\\
%does not  manifest itself in  the data}\\
  \vspace{1cm}
%Keywords:  large rapidity gap, diffractive dissociation, proton-nucleus collision, photon exchange\\
 %{\it arXiv: 2307.08625}\\

 \vfill

 E-mail: \url{v.a.khoze@durham.ac.uk}, 
  \url{ryskin@thd.pnpi.spb.ru}
 
\end{center}
 \newpage
\section{Introduction}
At high energies the processes of incoming hadron diffractive dissociation
constitute a significant part of the total cross-section. Its understanding is
quite important for Cosmic Ray physics since the formation of Extensive Air Showers is driven by the leading particles that carry a noticeable
fraction of initial hadron energy. The characteristic feature of such processes is the presence of a Large Rapidity Gap (LRG), which separates the
fast particles originated by hadron dissociation from the other secondaries.

Recently the CMS collaboration measured the rapidity gap distribution in proton-lead collisions at $\sqrt{s_{pN}}=8.16$ TeV~\cite{CMS-Pb}. For large rapidity gaps $\Delta\eta^F>2.5$, the distribution is practically flat ($d\sigma/d\Delta\eta^F\sim const$). However,
the popular Monte Carlo generators,  EPOS-LHC, QGSJET II, and HIJING,  underestimate 
the obtained cross-section by a factor of more than 2 for the case of the lead ion dissociation and more than a factor of 5 for the proton dissociation (note, however, that a very strong contribution
from $\gamma p$ interactions is not included in these event generators).\\

Recall that already for the 'elementary' proton-nucleon collisions the 
theoretical predictions for diffractive dissociation cross sections strongly depend on the value and the transverse momentum behavior of the triple-Pomeron, $g_{3P}$, vertex
 and multi($n\to m$)-Pomeron vertices, $g^m_n$ (see e.g.~\cite{gmn}), which are poorly known experimentally.
  Besides this, some nuclear effects are originating from the small
   'gap survival' probability, that is, the probability not to fill the LRG by
    the secondaries produced in an additional inelastic interaction of
     incoming proton with another lead ion nucleon.\\

In the present paper, we  focus on the nuclear effects 
%of extracting
% taking
tuning the parameters of the triple- and the multi-Pomeron vertices in such a way as to reproduce  
 the 'elementary' LRG cross-section 
from another CMS paper~\cite{CMS-7} where the value of  $d\sigma(pp)/d\Delta\eta^F$
 was measured at $\sqrt{s_{pp}}=7$ TeV.

Of course, the energy is not the same (7 and not 8.16 TeV).
Moreover, the event selection in \cite{CMS-7} differs from that in \cite{CMS-Pb}.
 In particular, in the proton-lead case of \cite{CMS-Pb}, we deal with the asymmetric geometry -- in the laboratory frame, the energy of the proton beam was 6.5 TeV while the nucleon in lead ion has the energy $6.5\cdot 82/208\simeq 2.56$ TeV. However, these differences could change the cross section
  by about 20 \% or so and cannot explain the factor of 2 (or 5) effect.

In section 2 we describe the event selection in the CMS experiment. Note that the cross-section 
 $d\sigma(pp)/d\Delta\eta^F$ includes not only the single dissociation (SD) of  one incoming
  hadron but also the double dissociation (DD) of both beam and target hadrons with the LRG in-between.

In section 3 we discuss the role of an additional interaction 
of the incoming proton with the nucleons in the lead.
The secondaries produced in these interactions fill the LRG, and this way reduces the observed number 
of 'gap' events.
 Actually,
 the LRG survives only in very peripheral collisions when the proton interacts with the relatively thin    
 edge of the heavy ion disk. In such a case, the probability of an additional interaction is sufficiently small.
 
 For this reason, the LRG cross-section is proportional to the 
 thickness of the disk edge.  It should be noted that here
we have to account not only for the width of the nucleon distribution in a heavy ion
 but for the radius of the proton-nucleon interaction as well. 
 At $\sqrt s=8.16$ TeV this radius is rather large (the elastic $t$ slope $B_{el}\simeq 20$ Gev$^{-2}$).  Accounting for the $B_{el}\neq 0$ in the additional/underlying inelastic 
 interactions could change the result up to 20\%.
  Even a stronger (up to a factor of 2) effect can be observed when
   we account 
   for the impact parameter structure of the 'elementary' $pN$ amplitude with the LRG.
    The structure of the elementary $pN$ amplitude with LRG in the transverse plane
     is calculated in Appendix A.

Next, in the case of the proton dissociation with a small momentum transferred, $q_T$, 
less than the inverse ion radius ($1/R$), the nucleons placed in the ion at the same impact parameter could act coherently.  This leads to the coherent contribution, $\sigma_{coh}$, more or less of the same order 
as the incoherent piece.

Section 4 is devoted to the photon exchange interaction with the lead ion,
%contribution, 
  which in the case of the proton dissociation
% interacting with the lead ion 
 turns out to be about 3 times larger than the incoherent dissociation caused by the Pomeron exchange.

We discuss the results in section 5 and conclude in section 6.

\section{LRG event selection}
To select the events with  LRG, the CMS collaboration used the following criteria.  Let us say that the proton beam has a positive rapidity while the lead beam has a negative rapidity. Then for the case of lead ion dissociation, the  forward calorimeter HF-   
\footnote{Forward hadron calorimeters HF$\pm$ cover the pseudorapidity region $ 2.85 < | \eta | < 5.19$ with the sign indicating the sign of $\eta$. } should have at
least one tower with energy greater than 10 GeV, and if in the central detector the bin $2.5 < \eta < 3$ is empty, the event is identified as a $\p Pb$ candidate and $\Delta\eta^F$ is defined as the distance from $\eta = 3$ to the nearest edge of the first nonempty $\eta$ bin. Alternatively, if 
  HF+
 has at least one tower above the threshold, and the bin 
$-3<\eta <-2.5$ is empty then the event is flagged as a $\p p+\gamma p$ candidate corresponding to the proton dissociation. (Here, by $\p$, we denote the Pomeron, which describes the interaction across the LRG.)  In this
case, $\Delta\eta^F$ is the distance from $\eta=-3$ to the nearest edge of the last nonempty $\eta$ bin.

Besides this, when presenting the final result an additional condition was added -- the calorimeter neighboring to the empty bin (say, HF- if the bin $-3<\eta <-2.5$ is empty) should have no detectable particles. That is, actually the LRG is started not from $|\eta_0|=3$ but from $|\eta_0|=5.19$. Nevertheless when presenting the results authors of \cite{CMS-Pb}
 still define the value of the gap size $\Delta\eta^F$ starting from $|\eta|=3$.
\begin{figure} [t]
\label{fig:dd}
\vspace{-5cm}
\hspace{-0.6cm}
        \includegraphics[scale=0.36]{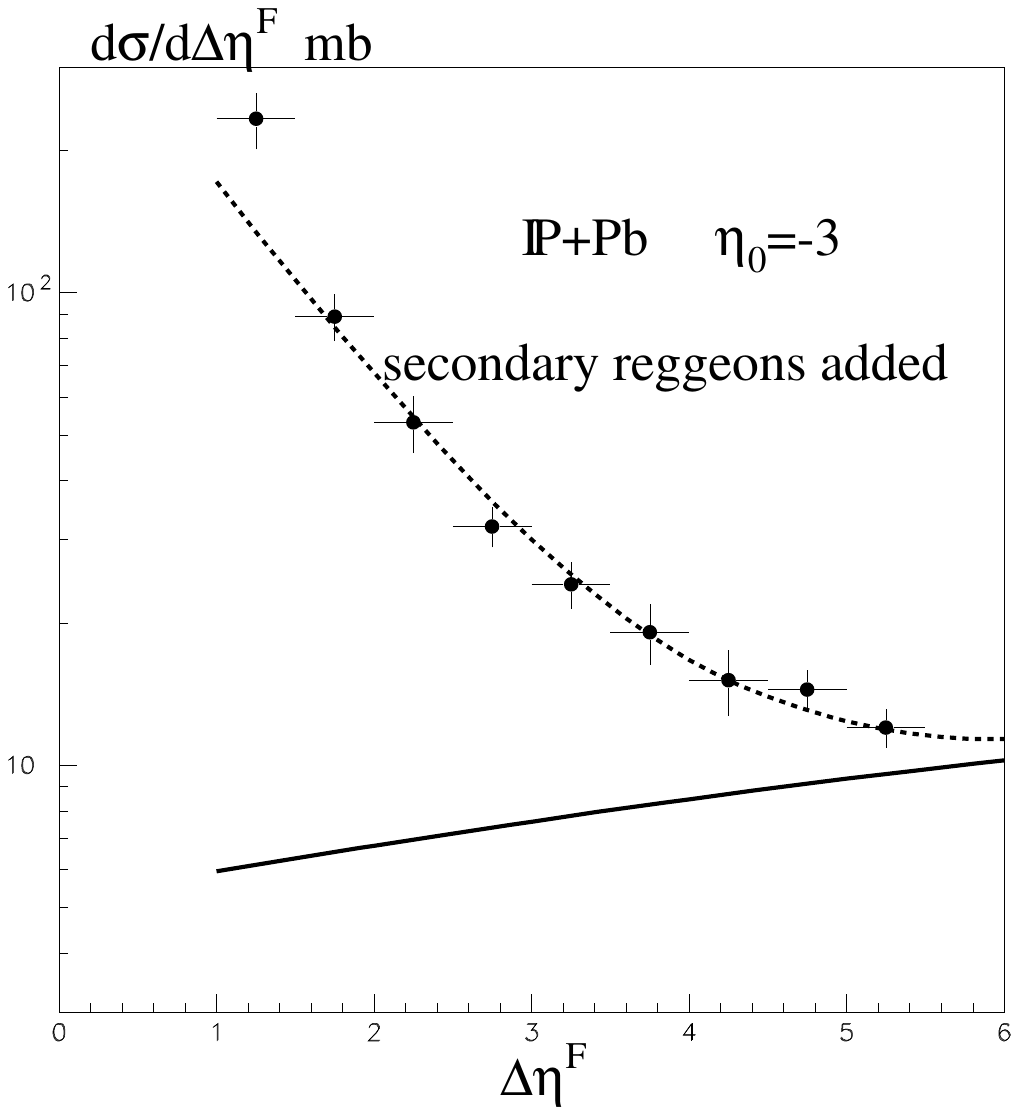}
\hspace{-1.2cm}
 \includegraphics[scale=0.36]{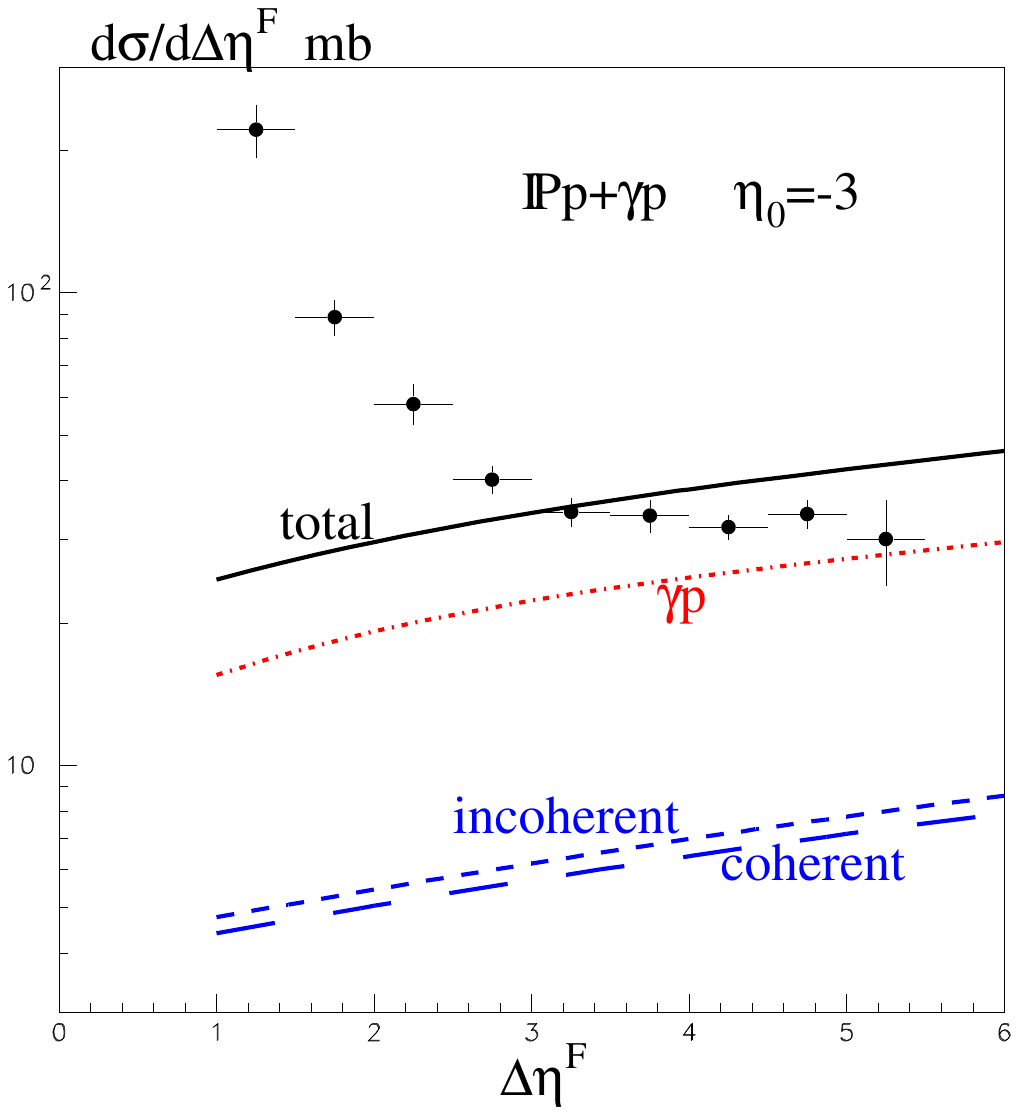}
\begin{center}
\vspace{-.4cm}
\caption{\small Differential cross section $d\sigma/d\Delta\eta^F$ for events with $\p Pb$ (left) and $\p p+\gamma p$ (right) topologies. The data are taken from \cite{CMS-Pb}. Solid black curves show the total cross sections while the short (long) dashed curves in the right panel are the incoherent (coherent) Pomeron contributions. Photon contribution is plotted by the dot-dashed red curve. A short dashed black curve in the left panel demonstrates the possible role of the secondary reggeons not included in the present model.}  
\label{f3}
\end{center}
\end{figure}
\begin{figure} [t]
\label{fig:dd}
\vspace{-5cm}
\hspace{-0.6cm}
        \includegraphics[scale=0.36]{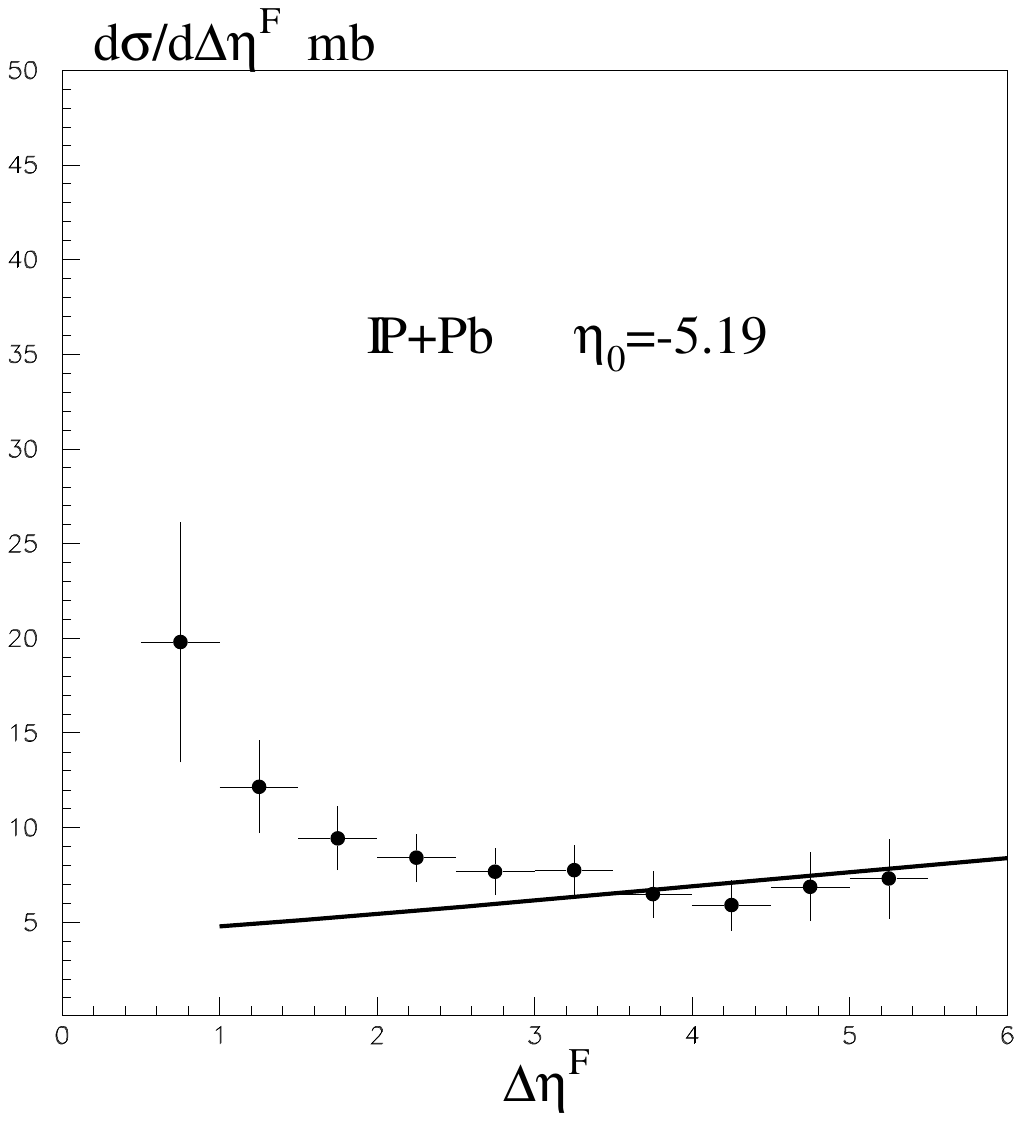}
\hspace{-1.2cm}
 \includegraphics[scale=0.36]{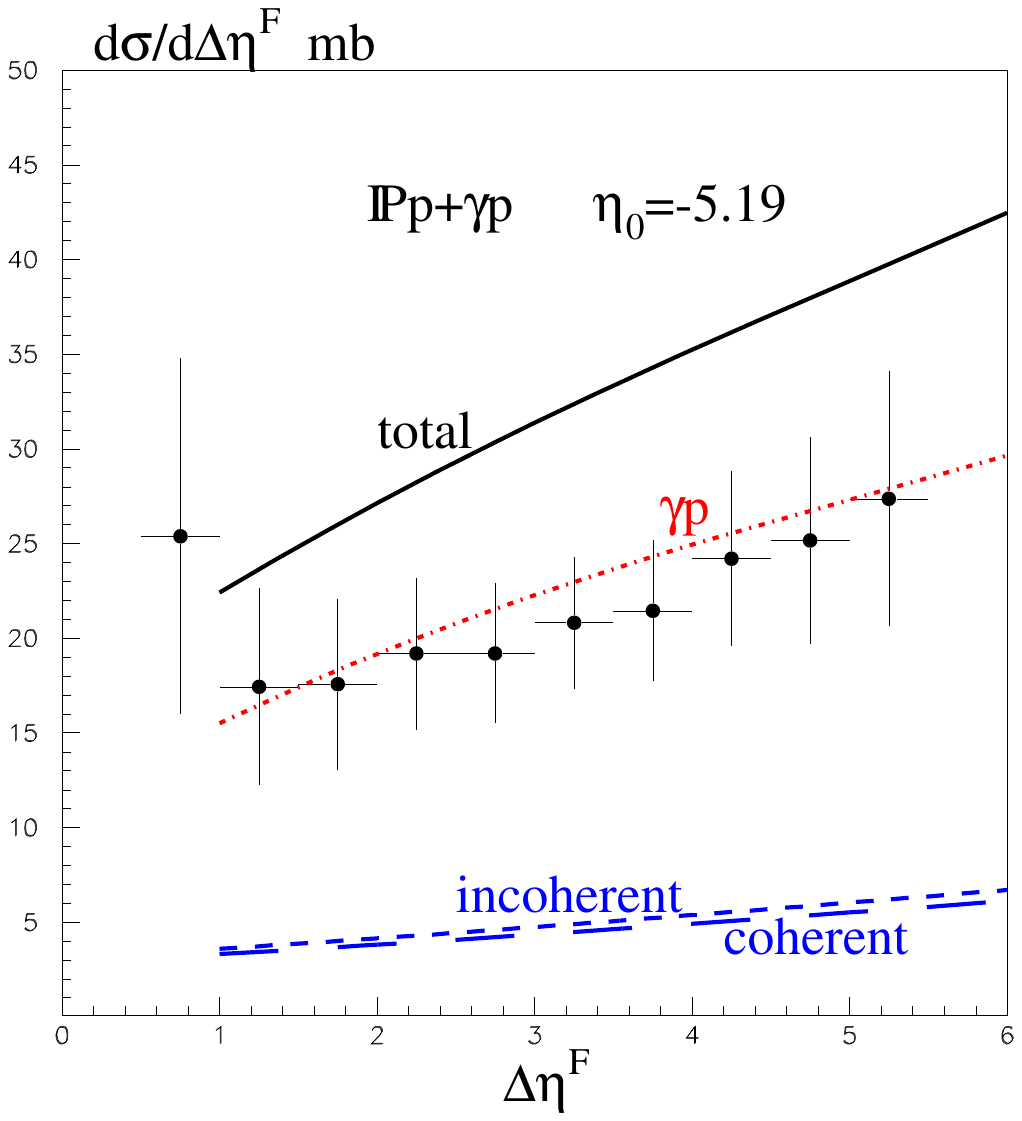}
\begin{center}
\vspace{-.4cm}
\caption{\small Differential cross section $d\sigma/d\Delta\eta^F$ for events with $\p Pb$ (left) and $\p p+\gamma p$ (right) topologies. The data are taken from \cite{CMS-Pb}.
Note that here the LRG starts not from $\eta_0=-3$ but from 
$\eta_0=-5.19$. That is actually the gap size is equal to $\Delta\eta^F+2.19$. 
 Solid black curves show the total cross sections while the short (long) dashed curves in the right panel are the incoherent (coherent) Pomeron contributions. Photon contribution is plotted by the dot-dashed red curve.}  
\label{f5}
\end{center}
\end{figure}

For the case without the forward calorimeter (HF) veto corresponding to the gap starting from $|\eta_0|=3$ our predictions are shown in Fig.\ref{f3}; both for the $\p Pb$ (left panel) and $\p p+\gamma p$ (right panel) configurations. 

As it is seen in Fig\ref{f3} the model strongly underestimates the cross section at relatively small $\Delta\eta^F$. Recall however that the model does not include the secondary reggeons. Naively the secondary reggeon contribution should decrease with $\Delta\eta^F$ as $\exp(-\Delta\eta^F)$. For this reason in the left panel we added the term $450\exp(-\Delta\eta^F)$ mb. The result is shown in the left panel by the dashed black curve.

The predictions corresponding to the gap starting from $|\eta_0|=5.19$ are shown in Fig.\ref{f5}.\\
Note that such a definition includes both the events with the only lead (or proton) dissociation (SD) (while the interaction on the opposite side is 'elastic') and the double (both sides) dissociation (DD) where all the 'opposite side' secondaries have the rapidity $|\eta|>5.19$.

Recall that all rapidities here are measured in the laboratory (and not the centre of mass) frame. On the other hand,
in the case of proton-lead collision, the elementary proton-nucleon interaction is asymmetric. The proton momentum $p=6.5$ TeV, while the momentum of the nucleon in lead ion is 2.56 TeV. Therefore the theoretical predictions for the values of
$d\sigma(pN)/d\Delta\eta^F$ in elementary proton-nucleon collisions are different for $\p Pb$ and the $\p p$ configurations.  The expected elementary cross-section in the $\p Pb$ (i.e., $\p N$) case becomes about 20\% larger.

\section{Gap survival factor}
Here we recall the structure of calculations needed to account for the gap survival probability in {\em exclusive} or  LRG events in the $p$Pb collisions. We  follow section 6 of~\cite{Lkmr} (see also~\cite{SC3}).

It is convenient to work in terms of the transverse coordinate, that is, using the impact parameters $b$.
The incoherent cross-section
\begin{equation}
\label{sig-inc}
\sigma_{\rm incoh}=\int d^2bT_e(b)|A|^2S^2(b)
\ ,
\end{equation}
where  $A=A(b_a)$ is the LRG proton-nucleon amplitude in the $b$ representation, and $S^2(b)$ is the gap survival factor. Taking $T_e(b)=T(b)$ equal to % that of the 
the density profile of 
lead ion, $T(b)$, we have to say that this equation is written in the limit of a small size amplitude $A$, i.e., the size of the amplitude is much less than that for the heavy ion, $R_{\rm Pb}$.
 However to account for the non-locality of $A$ in the computations we replace the  density profile $T(b)$ (\ref{T-opt}) by the convolution 
 $$T_e(b)=
 %\frac 1{4\pi B_V}
 \frac{\int d^2 b'\ T(b')|A(\vec b-\vec b')|^2}{\int d^2b' |A(b')|^2}\ .
 %\exp(-|\vec b-\vec b'|^2/4B_V)\ 
 $$
  Analogous replacements (with the corresponding slope $B_{\rm el}$) were used in the calculation of the survival factors $S^2$.
  $$T_s(b)=\frac 1{2\pi B_{el}}\int d^2 b'\ T(b')\exp(-|\vec b-\vec b'|^2/2B_{el})\ .$$
   For proton-nucleon interactions the slope $B_{el}=20$ GeV$^{-2}$ was used.   The density profile of the Pb ion, $T(b)$, is evaluated as given in Appendix B. The survival factor is
\begin{equation}
\label{gap-p}
S^2(b)=\exp\left(- <\nu(b)>\right),
\end{equation}  
which is the probability of not filling the gap by the secondaries produced in additional proton~-~lead interactions.  
\be
\label{nu}
<\nu(b)>=T_s(b)\sigma(pN)-1
\ee
 is the mean number of additional interactions of incoming proton
with the nucleons from the ion. We reduce the number of interactions by 1 since the first 'elementary' interaction with the LRG has already occurred.
Note that in (\ref{nu}) we have to account only for the inelastic interactions which fill the gap. That is
\be
\label{sig-gap}
\sigma(pN)=\sigma_{tot}-\sigma_{el} -\sigma_{LRG}\ .
\ee
In computation for $\sqrt s=8.16$ TeV we put $\sigma(pN)=65$ mb.\\

In the case of proton dissociation ($\p p$) when the momentum transferred to the lead ion is sufficiently small, the nucleons in the ion may act coherently.
For coherent interaction we have
\begin{equation}
\label{sig-coh}
\sigma_{\rm coh}=4\pi B_{sd}F^2_{\rm Pb}(t_{\rm min})\int  d^2b  
T^2_e(b)|A|^2S^2(b)\ ,
%S^2_V(b_V)\ ,
\end{equation}
where the dimension of extra $T_e$ factor is compensated by the $t$-slope of the elementary amplitude of dissociation \footnote{The $t$ behavior of dissociation cross section in the proton-nucleon collision is parameterized as $d\sigma^{SD}/dt\propto \exp(B_{sd}t)$ }, $B_{sd}=1/<k^2_T>$, where $<k^2_T>$ is the mean transverse momentum of the leading nucleon in elementary SD cross-section.
%The amplitude $A$ is normalized to $\int d^2b|A(b)|^2=\sigma(\gamma+p\to V+p)$. 
% The variables are sketched in Fig.4. 

The form factor $F_{\rm Pb}$ in (\ref{sig-coh}) accounts for the
nucleon distribution in the lead ion. The point is that the coherence of interaction with different nucleons should not be destroyed by the longitudinal component of the momentum transfer. This component is given by $t_{\rm min}=-(xm_N)^2/(1-x)$ where $x$ is the nucleon momentum fraction transferred across the gap.
% (by the Pomeron ?).
 Since the value of $|t_{\rm min}|$ is small in our kinematics, here we use just the exponential parametrization $F_{\rm Pb}(t)={\rm exp}(t\langle r^2_{\rm Pb}\rangle /6)$ with $\langle r^2\rangle $ being the mean radius squared of the lead ion.\\
\section{Photon exchange}
The probability of proton dissociation caused by the photon
radiated off the lead ion could be quite large since all the 82 protons in the $Pb$ act coherently, and the corresponding contribution is enhanced by the factor $Z^2=82^2=6724$.

For the amplitude with photon radiated off the lead ion, we consider only the coherent radiation, and since the photon transverse momentum here is very small, we keep just the 'electric' ($F_E$) term.
\be
\label{sig-Pb}
\frac{x\sigma(Pb\to\gamma +p)}{dx}=\sigma(\gamma+p)\int d^2b \frac{xd^3n_\gamma(x,b)}{dxd^2b}\cdot S^2(b)\ .
\ee 

The photon flux in the $b$ representation outside the heavy ion takes the form~\cite{Bud, best}
\be
\frac{d^3n_\gamma}{dxd^2b_\gamma} ~=~ \frac{Z^2\alpha^{\rm QED}}{x\pi^2 b_\gamma^2}~ (xm_Nb_\gamma)^2~K_1^2(xm_N b_\gamma),
\label{eq:fluxb}
\ee
where $K_1(z)$ is the modified Bessel function. \\
When the impact parameter $b$ becomes smaller than the ion size, the charge $Z$ should be replaced by the number of protons inside the sphere of radius $b$. However, this practically does not change the result since the  low $b$ (inside the ion) contribution is strongly suppressed by the gap survival factor $S^2(b)$.\\

For the $\gamma p$ cross-section the Donnachie-Landshoff parameterization~\cite{DL} is used
\be
\sigma(\gamma+p)=0.0677mb\cdot s^{0.0808}_{\gamma p}+0.129mb\cdot s^{-0.4525}_{\gamma p}\ ,
\ee
where the $\gamma p$ energy square $s_{\gamma p}$ is in GeV$^2$.

Since the position of LRG is known in terms of $\eta$, to evaluate the $\gamma p$ energy we assume that at the edge of the gap the mean transverse momentum of the pion, $<p_t>=0.7$ GeV.
Note that in the region of interest $s_{\gamma p}\sim 100-1000$ GeV$^2$ the variations of $\sigma(\gamma+p)$ value are rather small. So the exact value of $<p_t>$ is not crucial.

\section{Discussion}
In order not to worry about the secondary Reggeon contribution we consider
 the case with the forward calorimeter veto (Fig.\ref{f5}) and the largest (in~\cite{CMS-Pb}) rapidity gap $\Delta\eta^F=6$. For the 
 lead dissociation topology, ($\p Pb$), the situation looks quite reasonable. Starting with the elementary
 LRG cross-section $d\sigma(pN)/d\Delta\eta^F$ of about 0.5 mb (following~\cite{CMS-7})\footnote{ Our model gives 
 %elementary LRG cross-section 
 0.48 mb %for the $\p Pb$ case and 0.38 mb for the incoherent $\p p$ contribution
 .} we obtain for the $pPb$ case
$d\sigma/d\Delta\eta^F=8.4$ mb in agreement with~\cite{CMS-Pb}. %(8.2 mb for $\sigma^{abs}=70$ mb)

Note that neglecting the transverse size of the elementary amplitudes both for the LRG amplitude and the absorptive cross-section we get almost twice the smaller cross-section -- 4.6 mb instead of the 8.4 mb.\\

Recall that due to the small gap survival probability in the central (small $b$) collision already the 
elementary LRG amplitude is strongly suppressed at $b\to 0$. That is, the LRG amplitude is peripheral. An interesting fact is that practically the same (within 1\%) value of the LRG proton-lead cross section is obtained by replacing the $b$ shape of elementary amplitude given by our model by the simple,  $A\propto b\cdot\exp(-b^2/4B_{el})$, form where $B_{el}$ is the slope of elastic $pN$ cross-section.\\

The case of proton dissociation ($\p p+\gamma p$) looks more complicated. 
Due to asymmetry between the proton beam energy and the nucleon's momentum in the lead ion, the elementary cross section
 is expected to be about 20\%  smaller. %In our model, we get $d\sigma(pN)/d\Delta\eta^F=0.38$ mb instead of 0.48 mb. Therefore, here we start with $d\sigma(pN)/d\Delta\eta^F=0.4$ mb. 
 However, now we have to account for both the coherent ($d\sigma_{coh}/d\Delta\eta^F=6.1$ mb
 % (5.6 FOR $\sigma^{abs}=70$ mb)
  and incoherent $\p p$  contributions ($d\sigma_{incoh}/d\Delta\eta^F=6.7$ mb. %(6.564 for 70)).

Besides this, there is a large probability of destroying the beam proton by the photon coherently emitted off the ion. The expected cross-section is $d\sigma_\gamma/d\Delta\eta^F=29.6$ mb.\footnote{If we take $<p_t>=1$ GeV then $d\sigma_\gamma/d\Delta\eta^F=28.8$ mb.}
 This $\gamma p$ contribution is already equal to the cross-section observed by the CMS for  $\p p+\gamma p$ configuration. That is, we have no room for the $\p p$ contribution equal to 6.7+6.1=12.8 mb.
\subsection{Uncertainties}
Of course, the model used in this calculation has large uncertainties. The values and the $k_t$ behaviour of the multi-Pomeron vertices are not well known. On the other hand, we tuned the parameters to reproduce the CMS LRG results measured in proton-proton collisions~\cite{CMS-7}. Another argument in favour of a reasonable accuracy of our prediction is the fact that for the $\p Pb$ topology (Fig.\ref{f5}, left) the data at $\Delta\eta^F>3$ are described rather well. The main uncertainty of the photon contribution is the value of $\gamma$-proton cross-section where the error bars and the spread of the experimental data do not exceed 2-3\% (see \cite{DL}).  That is less than 1 mb for our case.

Finally, the result depends on the value of absorptive cross-section $\sigma(pN)$ used in (\ref{gap-p}) and (\ref{nu}) to calculate the gap survival probability. To estimate the possible effect we have replaced the value of $\sigma(pN)=65$ mb by $\sigma(pN)=70$ mb. For the events with the forward calorimeter veto (Fig.\ref{f5}) at $\Delta\eta^F=6$ this changes the total value of $d\sigma/d\Delta\eta^F$ from  42.5 mb to 41.2 mb for the $\p p+\gamma p$ topology and from  8.4 mb to 7.8 mb for the $\p Pb$ topology.

\subsection{Comparison with  Ref.\cite{ZS}}
Coherent contribution to the LRG proton-lead cross section in the
 ($\p p+\gamma p$) configuration was calculated in \cite{ZS}. For the $\gamma p$ channel the
 authors got practically the same result as that in this paper. However, they predicted a much smaller Pomeron-induced cross-section 
 $d\sigma_{coh}/d\Delta\eta^F=2.4\pm 1.3$ mb. So low value allowed them to state that the theory well reproduces the CMS measurement.
 
 Note that the incoherent contribution and the $\p Pb$ configuration were not considered in \cite{ZS}. On the other hand, we should expect the incoherent $\p p$ cross-section to be of the same order as that in the $\p Pb$ case (see e.g. MC predictions shown in Fig. 5 of  \cite{CMS-Pb}). This gives an additional 6-9 mb. 
 
 Besides this, the value of $d\sigma_{coh}/d\Delta\eta^F=2.4\pm 1.3$ mb was underestimated.\\
  First, they did not account for the size of the 'elementary' proton nucleon amplitude used in eq.(12) of \cite{ZS}, just the density profile of the lead ion. Indeed, the size of the elementary amplitude is much smaller than the lead ion radius, but it is larger than the width of the diffusion edge ($d\simeq 0.5$ fm) of the ion. In our case, the cross-section is proportional to the diffusion edge width. \\
  Next, the total $pp$ cross section, $<\sigma>=\sigma_{tot}=98.6$ mb, was 
  used in eq.(12) as the absorptive cross-section. However, in the CMS experiment, it was allowed to break the ion. That is, only the processes that fill the LRG act as absorptive corrections. Correspondingly one has to take a lower $<\sigma>\sim 60-70$ mb.
  
  As a result, the $\p p$ contribution is expected to be much larger, at about 10 mb, and it looks strange that in the case of incoming proton dissociation CMS  observed
the cross-section, which is well described by the pure photon
exchange.

\section{Conclusion}
The nuclear effects in the cross-section of the LRG events, 
$d\sigma/d\Delta\eta^F$, 
 measured by the CMS collaboration in the proton-lead collisions at 
$\sqrt s_{pN}=8.16$ TeV are discussed. It is shown that the size of the elementary $pN$ amplitude can not be neglected since it exceeds the width of the peripheral edge of the lead nucleus, $d\simeq 0.5$ fm. Accounting for the size of both the absorptive amplitude and the amplitude of LRG events leads 
to about twice the larger cross-section, which becomes consistent with that claimed by the CMS for the $\p Pb$ case.

The probability of the proton dissociation caused by the photon coherently emitted off the lead ion is so large that it can completely explain the whole LRG cross section observed by the CMS ($\sim 29$ mb) for the $\p p + \gamma p$ configuration.
If no additional experimental cuts %\\
% {\bf(possibly overlooked by us ???)}\\
were imposed to select  this configuration, then we face the problem -- why is the Pomeron-induced $\p p$ coherent plus incoherent 
contribution, which is expected to be about 13 mb, not seen in the data?
\\
 \section*{Appendix A:  LRG in proton-nucleon ($pN$) events}
 %Proton-nucleon ($pN$) events with the LRG}

To evaluate the profile of elementary $pN$ amplitude with LRG, we extended the two channel\footnote{The proton wave function has two Good-Walker~\cite{GW} components.}  model used in \cite{El} to successfully describe the elastic proton-proton (and proton-antiproton) interactions at collider energies.
The model was supplemented with the triple-Pomeron,
 $g_{3P}=g^2_1$, and the multi-Pomeron,
 $g^m_n$  vertices to reproduce the high mass diffractive dissociation.
We keep all the previous parameters from~\cite{El} and parameterise the multi-Pomeron vertices as
\be\label{3P}
g^m_n(k^2_i,k^2_j)=\lambda_1 (\lambda_2 g_N)^{m+n-2}\Pi_{i=1}^m e^{-b_{3P}k^2_i}\Pi_{j=1}^n e^{-b_{3P}k^2_j}\ ,
\ee
 where $g_N$ is the Pomeron nucleon coupling taking from \cite{El}, $k^2_i,\ k^2_j>0$ are the transverse momenta 
squared transferred through the corresponding Pomeron, and we put the slope $b_{3P}=0.5$ GeV$^{-2}$. $\lambda_1=0.4$ and $\lambda_2=0.5$, that is 
 $g_{3P}=\lambda_1\lambda_2g_N=0.2g_N$.
 
 The constant $\lambda_1=dn/dy$ is the density of the intermediate partons,  $dn/dy$, in the rapidity space while $\lambda_2$ is the ratio $g_{parton}/g_N=\lambda_2$ of the Pomeron coupling to the parton at the edge of LRG, $g_{parton}$, to the Pomeron-nucleon coupling, $g_N$.

\begin{figure} [t]
%\label{fig:dd}
\vspace{-5cm}
\hspace{-0.3cm}
        \includegraphics[scale=0.36]{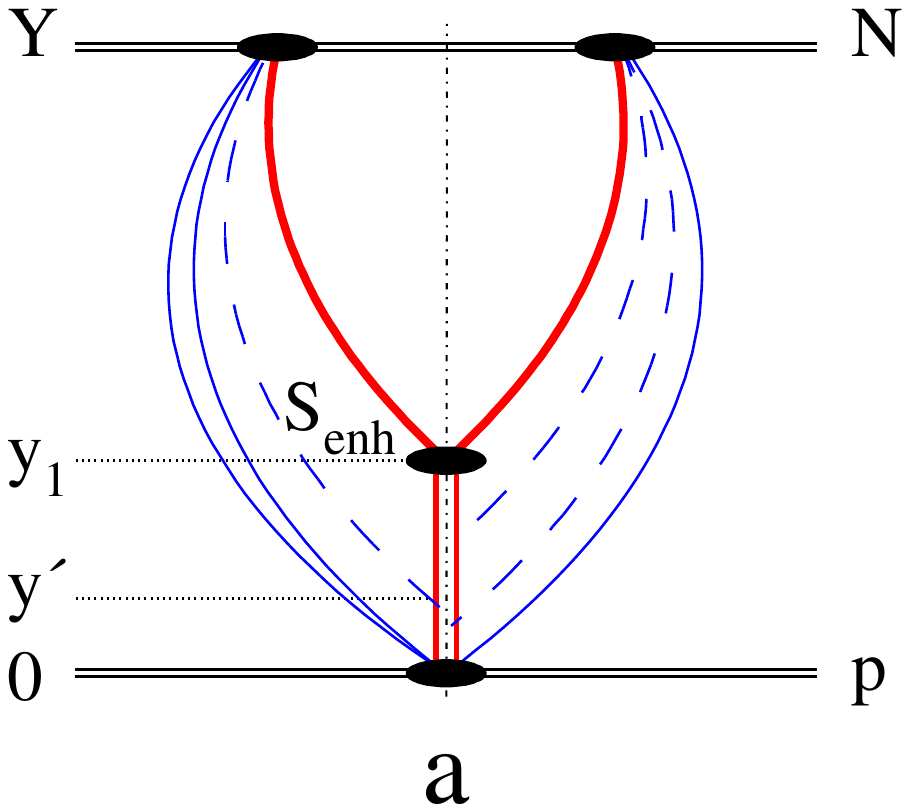}
\hspace{-1.2cm}
 \includegraphics[scale=0.36]{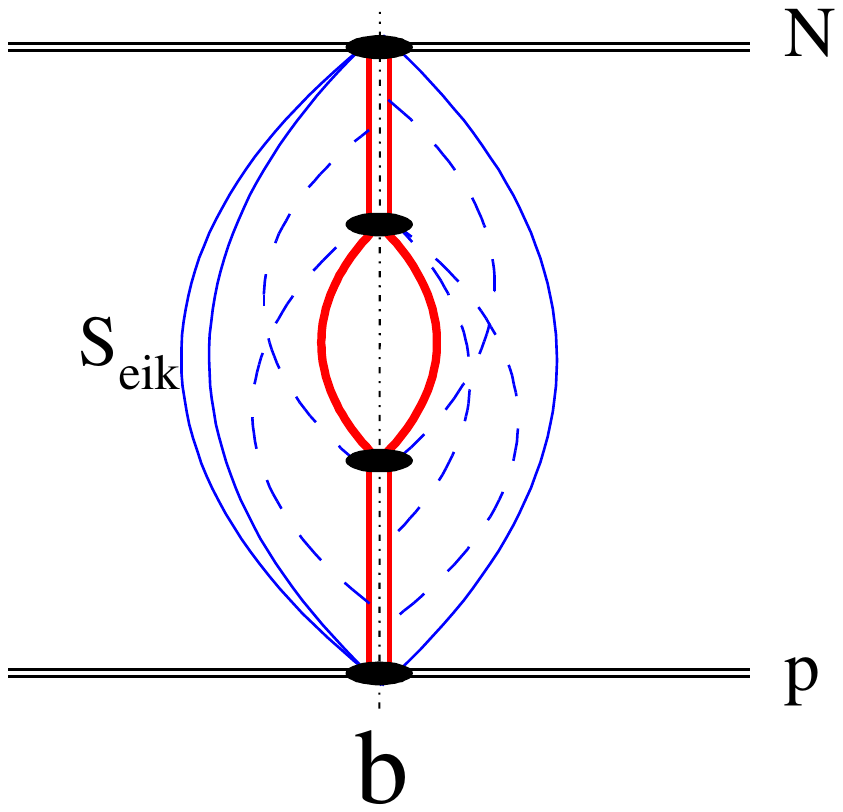}
\begin{center}
\vspace{-1.4cm}
\caption{\small Diagrams for the single (a) and double (b) diffractive dissociation. Heavy red lines denote the elastic amplitude $1-e^{-\Omega(\Delta b)/2}$ originated by the Pomeron(s) exchange across the LRG; double red lines cut by the dot-dashed line denote the inelastic Pomeron-induced amplitudes $1-e^{-\Omega(b)}$. In both cases, $\Omega(b)$ is the corresponding opacity caused by one Pomeron exchange in the $b$ representation. 
Dashed lines describe the semi-enhanced gap survival factor. Here we account for the whole inelastic interactions. That is, the dashed lines correspond to $(1-e^{-\Omega})$ and integration over the interactions with any intermediate Parton in central pomeron (from $y'=0$ to $y'=y_1$ in the case of the left figure and analogous for the upper part of the right figure) is included. The continuous thin blue lines indicate the non-enhanced Pomeron exchange. Its sum forms the eikonal survival factor $S^2_{eik}=\exp(-\Omega(b))$.}  
\label{f1}
\end{center}
\end{figure}

The corresponding diagrams are shown in Fig.\ref{f1}. We account for 
the eikonal gap survival factor, $S_{eik}$, (continuous blue lines) caused by the inelastic proton-nucleon interactions
and for the semi-enhanced screening, $S_{enh}$, due to inelastic interactions of a proton (or neutron) with the intermediate partons (blue dashed lines).  For the case of $S_{enh}$, we integrate over the parton rapidity $y'$ within the  interval covered by the cut Pomerons (double red line in Fig.\ref{f1}). That is for the single dissociation 
Fig.\ref{f1}a
\be
S^2_{enh}(\delta b)=exp\left(-\lambda_1\int_0^{y_1}dy'(1-e^{\Omega(\delta b)})\right)\ .
\ee
 $S^2_{eik}(b)=\exp(-\Omega(b))$. \\
The parameters of the Pomeron exchange used to calculate the corresponding opacity $\Omega(b)$ are the same as that in \cite{El}.\\

%The constant $\lambda_1$ is the density of the intermediate partons,  $dn/dy$, in rapidity space while $\lambda_2$ is the ratio $g_{parton}/g_N=\lambda_2$ of the Pomeron coupling to the parton at the edge of LRG, $g_{parton}$, to the Pomeron-nucleon coupling, $g_N$.

%<\lambda_2$ since besides the ratio $g_{parton}/g_N=\lambda_2$ of the Pomeron coupling to the parton at the edge of LRG, $g_{parton}$ to the Pomeron-nucleon coupling, $g_N$ 
%it contains the density of the intermediate partons,
% $dn/dy$, in rapidity space. $\lambda_1=\lambda_2\cdot  dn/dy$.\\
 
 The model reasonably reproduces the existing (limited) data on LRG events cross sections at the LHC energies with the multi-Pomeron vertices described above. 
% {\bf Do we need here to put the reference to the expert? data?
%VAK  I think not, we can reference here, see \cite{El}}\\
% However, due to the rather large freedom in choosing the parameters in the present paper, we use this model just to calculate the impact parameter ($b$) structure of the LRG event amplitude, while
% for the absolute value of the LRG cross-section, we prefer to base directly on the experimental data. \\
 
\newpage
{\bf\Large Appendix B: 
Density profile of the Pb ion}

The  density profile of the Pb ion can be written in the form
\begin{equation}
\label{T-opt}
T(b)=\int_{-\infty}^{+\infty}dz(\rho_p(r)+\rho_n(r)) 
\end{equation}
with $r=\sqrt{z^2+r^2_t}$ and $r_t=b$. For the nucleon density in the lead $\rho(r)$ we use 
the Wood-Saxon form~\cite{Woods}
\be
\label{rho}
\rho_N(r)= \frac{\rho_0}{1+\exp{((r-R)/d)}}\;,
\ee
where the parameters  $d$ and $R$ respectively characterise the skin thickness and the radius of the nucleon density in the heavy ion; $r=(z,r_t)$. For $^{208}$Pb
we take the recent results of~\cite{Tarbert,Jones}
\begin{align}\nonumber
R_p &= 6.680\, {\rm fm}\;, &d_p &= 0.447 \, {\rm fm}\;,\\ \label{eq:pbpar}
R_n &= (6.67\pm 0.03)\, {\rm fm}\;, &d_n &= (0.55 \pm 0.01) \, {\rm fm}\;.
\end{align}
The nucleon densities, $\rho$, are normalized to 
\be
 \int\rho_p(r)d^3r=Z \;, \qquad \int\rho_n(r)d^3r=N_n\;,
\ee
for which the corresponding proton (neutron) densities are $\rho_0 = 0.063$ (0.093) ${\rm fm}^{-3}$.

  \section*{Acknowledgments}  
  The authors thank V.T. Kim and L. Kheyn for the discussion.
% and the reading of the manuscript.\\
  
%  \section*{Data Availability Statement} This manuscript has no associated data or the data will not be deposited. [Authors’ comment: We  discuss the data presented by CMS collaboration in Ref.\cite{CMS-Pb} and there are no other  data to be deposited to accompany this article.]
  
%\section*{Declarations}
%No funds, grants, or other support were received.


\thebibliography{}
\bibitem{CMS-Pb} CMS Collaboration, arXiv: 2301.07630.
\bibitem{gmn} S.S. Ostapchenko, Contribution to: 40th Rencontres de Moriond on QCD and High Energy Hadronic Interactions, 183-186, e-Print: hep-ph/0504164;\\
    M.G. Ryskin, A.D. Martin, V.A. Khoze,  Eur.Phys. J.C {\bf 60} (2009) 249,  e-Print: 0812.2407 [hep-ph].
\bibitem{CMS-7}     CMS Collaboration, Vardan Khachatryan 
et al.,  Phys.Rev.D {\bf 92} (2015) 1, 012003, 
arXiv: 1503.08689 [hep-ex]
\bibitem{Lkmr} %Searching for the Odderon in Ultraperipheral Proton--Ion Collisions at the LHC
L.A. Harland-Lang, V.A. Khoze, A.D. Martin, M.G. Ryskin, Phys.Rev.D 99 (2019) 3, 034011, e-Print:1811.12705 [hep-ph (sect.6)].
\bibitem{SC3}
L.A. Harland-Lang, V.A. Khoze, M.G. Ryskin,Eur.Phys. J.C 79 (2019) 1, 39; e-Print:1810.06567 [hep-ph].

\bibitem{Bud}  V. M. Budnev, I. F. Ginzburg, G. V. Meledin, and V. G. Serbo, Phys.Rept. {\bf 15}, 181 (1975).
\bibitem{best} M. Vidović, M. Greiner, C. Best, G. Soff,  %Impact-parameter dependence of the electromagnetic particle production in ultrarelativistic heavy-ion
%collisions, 
Phys. Rev. {\bf C47}, 2308 (1993).
\bibitem{DL}     A. Donnachie, P.V. Landshoff,  Phys.Lett. B {\bf 296} (1992) 227, e-Print: hep-ph/9209205 [hep-ph];     Acta Phys.Polon.B {\bf 27} (1996) 1767.

\bibitem{ZS} V. Guzey, M. Strikman, M. Zhalov, Phys. Rev. C {\bf 106}, L021901 (2022). 


\bibitem{GW}  M. L. Good and W. D. Walker,
% Diffraction disssociation of beam particles, 
Phys. Rev. {\bf 120} (1960) 1857.
\bibitem{El} V.A. Khoze, A.D. Martin and M.G. Ryskin, 
    Phys.Lett.B 784 (2018) 192, e-Print: 1806.05970 [hep-ph].

\bibitem{Woods} R.D. Woods and D.S. Saxon, Phys. Rev. {\bf 95} (1954) 577.


\bibitem{Tarbert} C.M. Tarbert et al., Phys. Rev. Lett. {\bf 112} (2014) 242582.

\bibitem{Jones} A.B. Jones and B.A. Brown, Phys. Rev. {\bf C98} (2014) 067384.
\end{document}